\documentclass[aps,preprint,floatfix,nofootinbib,showpacs]{revtex4-1}
\pdfoutput=1
\usepackage{graphicx,color,float}
\usepackage{hyperref}

\begin{document}


\title{
Measuring properties of a Heavy Higgs boson\\[0mm] in the 
$H\to ZZ \to 4\ell$ decay
}

\def\slash#1{#1\!\!/}
\def\lsim{\:\raisebox{-0.5ex}{$\stackrel{\textstyle<}{\sim}$}\:}
\def\gsim{\:\raisebox{-0.5ex}{$\stackrel{\textstyle>}{\sim}$}\:}
\newcommand{\imag}{\Im {\rm m}}
\newcommand{\real}{\Re {\rm e}}

\renewcommand{\thefootnote}{\arabic{footnote}}

\author{
Jung Chang$^{1,2}$, Kingman Cheung$^{2,3,4}$, Jae Sik Lee$^{1,2,5}$,
Chih-Ting Lu$^4$, and  Jubin Park$^{5,1,2}$}
\affiliation{
$^1$ Department of Physics, Chonnam National University, 
300 Yongbong-dong, Buk-gu, Gwangju, 500-757, Republic of Korea \\
$^2$ Physics Division, National Center for Theoretical Sciences,
Hsinchu, Taiwan \\
$^3$ Division of Quantum Phases and Devices, School of Physics, 
Konkuk University, Seoul 143-701, Republic of Korea \\
$^4$ Department of Physics, National Tsing Hua University,
Hsinchu 300, Taiwan \\
$^5$ Institute for Universe and Elementary Particles, Chonnam National University, \\
300 Yongbong-dong, Buk-gu, Gwangju, 500-757, Republic of Korea 
}
\date{November 5, 2017}

\begin{abstract}
In many extensions of the standard model, there exist a few extra
Higgs bosons. Suppose a heavy neutral Higgs boson $H$ is
discovered at the LHC, one could then investigate 
CP and 
CP$\widetilde{\rm T}$ properties of its couplings to a pair of $Z$ bosons
through $H \to ZZ \to 4\ell$. We use the helicity-amplitude method to
write down the most general form for the angular distributions
of the four final-state leptons, which
can cover the case of CP-even, -odd, and -mixed state for the Higgs
boson.  
We figure out there are 9 types of angular observables 
and all the $H$ couplings to $Z$ bosons
can be fully determined by exploiting them.
A Higgs-boson mass of 260 GeV below the $t\bar t$ threshold is
illustrated with full details.  With a total of $10^3$ events of $H
\to ZZ \to 4\ell$, one can determine the couplings up to 
12-20\% uncertainties.
\end{abstract}

\maketitle

\section{Introduction}
The measured properties of the scalar boson which was discovered
at the LHC~\cite{atlas,cms} turn out to be the best 
described by the Standard Model (SM) Higgs boson~\cite{Cheung:2013kla}
and it deserves to be called the Higgs boson
which was proposed in 1960s~\cite{higgs}.
Among the Higgs boson couplings to the SM particles,
the most constrained one is its coupling to the massive gauge bosons
normalized to the corresponding SM value:
$C_v = 0.94\,^{+0.11}_{-0.12}$.
\footnote{
For the reference value of the coupling $C_v$, 
we have taken the 1-$\sigma$ range 
obtained upon the LHC Run-1 data
by varying the Higgs couplings to
the top- and bottom-quarks, $\tau$ leptons, gluons, photons,
and the massive gauge bosons
under the assumption that the 125 GeV Higgs boson carries the CP-even 
parity~\cite{Cheung:2014noa}.
%
%
}

Even though the SM has achieved a great success in describing the 
interactions among the basic building blocks of matter scrutinized by now,
however more blocks and new interactions are required to explain
the experimental observations of dark matter, non-vanishing neutrino mass, 
the baryon asymmetry of our Universe, inflation, etc.
In most extensions beyond the SM,  the Higgs sector is enlarged to include more
than one Higgs doublet resulting in charged Higgs bosons and several 
neutral Higgs bosons in addition to the one discovered at the LHC.
For example, the minimal supersymmetric extension of the 
SM, aka MSSM~\cite{HPN}, requires two Higgs doublet fields, thus leading 
to a pair of charged Higgs bosons and 3 neutral ones. 
In the next-to-minimal supersymmetric standard model, there are 
two additional neutral Higgs bosons~\cite{Cheung:2010ba}.
As another example, the Higgs Triplet Model that can explain
the mass spectrum and mixing of neutrinos gives rise to a pair of
doubly-charged Higgs bosons, a pair of singly-charged Higgs bosons, and 
3 neutral ones~\cite{Akeroyd:2005gt}.

Suppose that in future experiments a neutral Higgs boson $H$
heavier than the SM 125 GeV Higgs boson (denoted by $h$) is discovered.
Below the decay threshold into a top-quark pair or when $M_H<2m_t$,
assuming $H$ does not carry any definite CP-parity,
it may mainly decay into a bottom-quark pair ($b\bar b$), 
tau leptons ($\tau^+\tau^-$),
massive vector bosons ($W^+W^-$ and $ZZ$), 
a pair of $125$ GeV Higgs bosons ($hh$),
and a massive gauge boson and a lighter Higgs boson ($hZ$).
Above the $2m_t$ threshold, the decay mode into a top-quark pair may dominate
as in the MSSM 
\footnote{
We refer to Ref.~\cite{Lee:2003nta}
and references therein  for the typical
decay patterns of the heavy MSSM neutral Higgs bosons which do not
carry any definite CP parities.}.

The fermionic decay modes of $H\to b\bar b, \tau^+\tau^-, t\bar t$ 
and one of the bosonic decay modes $H\to W^+W^-$ may suffer 
from large QCD backgrounds and/or missing neutrinos. 
Among the remaining
bosonic decay modes into $ZZ$, $hh$, and $hZ$, 
taking account of the spin-$0$ nature of $H$,
only the  $ZZ$ mode may lead to nontrivial angular correlations among 
the decay products of the $Z$ bosons through the interferences among
various helicity states of the two intermediate $Z$ bosons
before their decays.

In this work, we consider the decay $H\to ZZ$ with the $Z$ bosons subsequently
decaying into electrons and/or muons: $H\to ZZ\to 4\ell$.
Long before the discovery of the SM Higgs boson,
it was suggested to exploit this decay process to determine 
the spin and parity of the Higgs boson~\cite{Choi:2002jk}. 
Later, more rigorous angular analyses of spin-zero, -one, and -two resonances
were illustrated with certain levels of experimental 
simulations~\cite{Gao:2010qx}.
After the 125 GeV Higgs-boson discovery, the method was practically applied to
determine the spin and CP properties of 
the ``newly" discovered boson~\cite{Bolognesi:2012mm,Modak:2013sb}.
Here, we shift the focus from the SM Higgs to a heavy Higgs boson $H$
\footnote{
For a detailed analysis on a heavy spin 1 resonance, 
see Ref.~\cite{Modak:2014zca}.},
and pursue complete determination of its couplings
from the angular correlations among the charged leptons in the final state.
Under the current experimental status, in which active 
searches for heavy resonances decaying into a $ZZ$ pair 
have been continually performed~\cite{h_zz},
our study may show how well one can determine
the properties of such a heavy scalar Higgs boson
at the LHC and/or High Luminosity LHC (HL-LHC).

The remainder of this article is organized as follows. In Sec.~II,
based on the helicity amplitude method~\cite{Hagiwara:1985yu},
we present a formalism for the study of angular distributions
in the decay $H\to ZZ\to 4\ell$.
We point out that there can be 9 angular observables in general 
and we can classify them according to 
the CP and CP$\widetilde{\rm T}$ parities of each observable.
In Sec.~III, we illustrate how well one can measure the couplings of 
a heavy Higgs boson by exploiting the angular observables
introduced in Sec.~II.
Finally, Sec.~IV is devoted to a brief summary,
some prospects for future work and conclusions.

\newpage
\section{Formalism}
One may start by defining the interaction of the heavy Higgs boson $H$ 
with a pair of $Z$ bosons.  
The amplitude for the decay process $H \to Z(k_1,\epsilon_1)\ Z(k_2,\epsilon_2)$
can be written as
\footnote{
Throughout this paper, we use the following abbreviations:
$s_\theta\equiv\sin\theta$, $c_\theta\equiv\cos\theta$,
$s_\Phi\equiv\sin\Phi$, $c_\Phi\equiv\cos\Phi$,
$c_{2\Phi}\equiv\cos\,2\Phi$, 
$s_{2\Phi}\equiv\sin\,2\Phi$, 
$s_W\equiv\sin\theta_W$,
$c_W\equiv\cos\theta_W$, etc.
}
\begin{eqnarray}
i{\cal M}^{H\to ZZ}
&\equiv & i \frac{gM_W}{c_W^2}\
\Gamma^{ZZ}_{\mu\nu} \epsilon_1^{*\mu}\epsilon_2^{*\nu}
\nonumber \\[2mm]
&=& i\frac{g M_W}{c_W^2}\Bigg\{
g_{_{HZZ}}\, \epsilon_1^*\cdot\epsilon_2^*
+S^{ZZ}_H(s) \left[ \frac{-2k_1\cdot k_2}{s}\,\epsilon_1^*\cdot\epsilon_2^*
\ + \ \frac{2}{s}\,k_1\cdot\epsilon_2^*\,k_2\cdot\epsilon_1^* \right]
\nonumber \\[2mm]
&& \hspace{3.41cm}
+ \ P^{ZZ}_H(s)\, \frac{2}{s}\, \langle \epsilon_1^*\epsilon_2^* k_1 k_2\rangle
\Bigg\}
\end{eqnarray}
where $k_{1,2}$ and $\epsilon_{1,2}$ are the four-momenta 
and the wave vectors of the two $Z$ bosons, respectively, with $s=(k_1+k_2)^2=M_H^2$
and $\langle \epsilon_1^*\epsilon_2^* k_1 k_2\rangle \equiv 
\epsilon_{\mu\nu\rho\sigma}\epsilon_1^{*\mu}\epsilon_2^{*\nu} k_1^\rho k_2^\sigma$.
The first term may come from the dimension-four renormalizable operator
\begin{equation}
{\cal L}=\frac{gM_W}{2c_W^2}\ g_{_{HZZ}}\ Z_\mu Z^\mu H
\end{equation}
while the form factors $S^{ZZ}_H$ and $P^{ZZ}_H$
can be generated by including higher-order corrections and/or
introducing non-renormalizable operators.
In the former case, $S^{ZZ}_H$ and $P^{ZZ}_H$ can be complex
by developing non-vanishing absorptive parts
in the existence of (New Physics)  particles 
running in the loop with mass less than $M_H/2$.
Therefore, in general one may need 5 real parameters to describe 
the interaction of the heavy Higgs boson $H$ with a pair of
$Z$ bosons.  
Note that $g_{_{HZZ}}^2\leq 1-g_{_{hZZ}}^2=1 -C_v^2$ with equality 
holding when $h$ and $H$ are the only Higgs bosons 
participating in the electroweak-symmetry  breaking.
We observe that being different from the case of SM Higgs boson, in which 
$g_{_{hZZ}}$ is dominating over the loop-induced $S_h^{ZZ}$ and $P_h^{ZZ}$ 
couplings, 
each of the couplings $g_{_{HZZ}}$, $S_H^{ZZ}$, and $P_H^{ZZ}$ 
may contribute comparably in the heavy Higgs-boson case.
We further observe that either 
$g_{_{HZZ}}\times P_H^{ZZ}\neq 0$  or $S_H^{ZZ}\times P_H^{ZZ}\neq 0$
implies that $H$ is a CP-mixed state, thus signaling CP violation.

Incidentally, the interaction of the $Z$ boson with a fermion pair is described
by the interaction Lagrangian:
\begin{equation}
{\cal L}_{Z f f}=
-\frac{g}{c_W}\ \bar{f}\gamma_\mu (v_f-a_f\gamma_5)f\ Z^\mu
= -\frac{g}{c_W}\
\sum_{A=+(R),-(L)}
\bar{f}\gamma_\mu (v_f-A a_f) P_A f\ Z^\mu
\end{equation}
with $v_f=I_3^f/2-Q_fs_W^2$, $a_f=I_3^f/2$ and $P_A=(1+A\gamma_5)/2$.

\subsection{Helicity amplitude}
We first present the helicity amplitude for the process 
$H\ \to Z(k_1,\epsilon_1) Z(k_2,\epsilon_2) \to
f_1(p_1,\sigma_1) \bar{f}_1(\bar{p}_1,\bar\sigma_1) \
f_2(p_2,\sigma_2) \bar{f}_2(\bar{p}_2,\bar\sigma_2)$.
Here, $p_{1,2}$ and and $\bar{p}_{1,2}$ are
four-momenta of the fermions $f_{1,2}$ and $\bar{f}_{1,2}$, respectively,
with $k_{1,2}=p_{1,2}+\bar{p}_{1,2}$.
And we denote the helicities of $f_{1,2}$ and $\bar{f}_{1,2}$ by
$\sigma_{1,2}$ and $\bar\sigma_{1,2}$.
Depending on the helicities of the four final-state fermions,
the amplitude  can be cast into the form
\begin{eqnarray}
i{\cal M}_{\sigma_1\bar\sigma_1 : \sigma_2\bar\sigma_2}
&=& \left(i\frac{gM_W}{c_W^2}\Gamma^{ZZ}_{\mu\nu}\right)\
\frac{-i\left(g^{\mu\rho}-\frac{k_1^\mu k_1^\rho}{M_Z^2}\right)}
{k_1^2-M_Z^2+iM_Z\Gamma_Z}\
\frac{-i\left(g^{\nu\sigma}-\frac{k_2^\nu k_2^\sigma}{M_Z^2}\right)}
{k_2^2-M_Z^2+iM_Z\Gamma_Z} \nonumber \\[3mm]
&\times &
\left[-i\frac{g}{c_W}\ \sum_{A=L,R} \bar{u}(p_1,\sigma_1)\gamma_\rho
(v_{f_1}-Aa_{f_1}) P_A v(\bar p_1 ,\bar\sigma_1)\right] \nonumber \\[3mm]
&\times &
\left[-i\frac{g}{c_W}\ \sum_{B=L,R} \bar{u}(p_2,\sigma_2)\gamma_\sigma
(v_{f_2}-Ba_{f_2}) P_B v(\bar p_2 ,\bar\sigma_2)\right] \nonumber \\[4mm]
&=& i \sum_{\lambda_1,\lambda_2}{\cal M}^{H\to ZZ}_{\lambda_1\lambda_2}\
\frac{1} {k_1^2-M_Z^2+iM_Z\Gamma_Z}\
\frac{1} {k_2^2-M_Z^2+iM_Z\Gamma_Z} \
{\cal M}^{Z\to f_1\bar f_1}_{\lambda_1:\sigma_1\bar\sigma_1}
{\cal M}^{Z\to f_2\bar f_2}_{\lambda_2:\sigma_2\bar\sigma_2}\nonumber \\
\end{eqnarray}
using
\begin{equation}
-g_{\mu\nu}+\frac{k_\mu k_\nu}{m^2} =\sum_{\lambda=\pm,0}
\epsilon_\mu^*(k,\lambda)\epsilon_\nu(k,\lambda)\,.
\end{equation}
The helicity amplitude for the decay 
$H\ \to Z(k_1,\epsilon_1) Z(k_2,\epsilon_2)$
in the rest frame of $H$ is given by
\begin{equation}
{\cal M}^{H\to ZZ}_{\lambda_1\lambda_2}=\frac{gM_W}{c_W^2}\,
\langle\lambda_1\rangle\ \delta_{\lambda_1\lambda_2}
\end{equation}
with the reduced amplitudes $\langle\lambda_1\rangle$ defined by
\begin{eqnarray}
\label{eq:HZZ}
\langle+\rangle&\equiv & - g_{_{HZZ}}
+(1-\alpha_1-\alpha_2)\, S^{ZZ}_H\ -i \lambda^{1/2}(1,\alpha_1,\alpha_2)\, P^{ZZ}_H\,,
\nonumber \\[2mm]
\langle-\rangle&\equiv & - g_{_{HZZ}}
+(1-\alpha_1-\alpha_2)\, S^{ZZ}_H\ +i \lambda^{1/2}(1,\alpha_1,\alpha_2)\, P^{ZZ}_H\,,
\nonumber \\[2mm]
\langle0\rangle&\equiv &g_{_{HZZ}}\left(
\frac{1-\alpha_1-\alpha_2}{2\sqrt{\alpha_1\alpha_2}}\right)
-2\sqrt{\alpha_1\alpha_2}\, S^{ZZ}_H\,,
\end{eqnarray}
where $\lambda(x,y,z) = x^2+y^2+z^2-2xy-2yz-2zx$ and $\alpha_i=k_i^2/M_H^2$.
We note that the contribution of $g_{_{HZZ}}$ to the longitudinal amplitude
$\langle0\rangle$ is enhanced by a factor $M_H^2/2M_Z^2$ in the large $M_H$ limit.

On the other hand,
the helicity amplitude for the decay 
$Z(k,\epsilon(k,\lambda)) \to f(p,\sigma)\bar f(\bar{p},\bar\sigma)$ 
is given by
\begin{eqnarray}
{\cal M}^{Z\to f\bar f}_{\lambda : \sigma\bar\sigma}=
\left\{ \begin{array}{ll}
-\frac{g}{c_W}\left[
\sqrt{2} m_f v_f\ \lambda\sigma e^{-i(\sigma-\lambda) \phi}\ s_\theta\
\delta_{\sigma\bar\sigma}\right.  &  \\[4mm]
\left.
\hspace{0.8cm}
+\frac{\sqrt{k^2}}{\sqrt{2}}(v_f-\sigma\beta_f a_f)(\lambda c_\theta+\sigma)\
e^{i\lambda \phi}\ \delta_{\sigma-\bar\sigma}\right]
& {\rm for}~\lambda=\pm  \\[4mm]
-\frac{g}{c_W}\left[
2 m_f v_f\ e^{-i\sigma \phi}\ (-\sigma c_\theta)\
\delta_{\sigma\bar\sigma}
+\sqrt{k^2} (v_f-\sigma\beta_f a_f) s_\theta\
\delta_{\sigma-\bar\sigma}\right]
& {\rm for}~\lambda=0
\end{array} \right.
\end{eqnarray}
in the rest frame of the fermion pair.
Note that the $Z$ boson is moving to the positive $z$ direction in
the $H$-rest frame, and $\theta$ and $\phi$ denote the polar and azimuthal
angles of the momentum $p$ of $f$ in fermion-pair rest frame.

Collecting all the sub-amplitudes and 
neglecting the masses of the final-state fermions,
we obtain
\begin{eqnarray}
\label{eq:hzz_helamp}
{\cal M}_{\sigma_1\bar\sigma_1 : \sigma_2\bar\sigma_2}
&=& \frac{gM_W}{2c_W^2}\left(\frac{g}{c_W}\right)^2\
\frac{\sqrt{k_1^2}} {k_1^2-M_Z^2+iM_Z\Gamma_Z}\
\frac{\sqrt{k_2^2}} {k_2^2-M_Z^2+iM_Z\Gamma_Z} \nonumber \\[3mm]
&\times &
(v_{f_1}-\sigma_1 a_{f_1}) (v_{f_2}-\sigma_2 a_{f_2}) \\[3mm]
&\times &
\left[
 \langle +\rangle (c_{\theta_1}+\sigma_1) (c_{\theta_2}+\sigma_2) e^{i(\phi_1+\phi_2)}
+\langle -\rangle (-c_{\theta_1}+\sigma_1)
(-c_{\theta_2}+\sigma_2) e^{-i(\phi_1+\phi_2)}\right.
\nonumber \\[3mm]
&&\hspace{0.0cm} \left.
+2\langle 0\rangle s_{\theta_1}s_{\theta_2} \right]
\delta_{\sigma_1-\bar\sigma_1} \delta_{\sigma_2-\bar\sigma_2}\,.
\nonumber
\end{eqnarray}
We observe the amplitude is receiving contributions from all the three helicity states 
$\langle+\rangle$, $\langle-\rangle$, and $\langle0\rangle$
of the intermediate $Z$ bosons, and the interferences among
the different helicity states lead to non-trivial angular distributions.

\subsection{Angular coefficients}
Neglecting the masses of the charged leptons in the final state, 
we find that the amplitude squared can be organized as:
\begin{eqnarray}
\sum_{\sigma_1,\bar\sigma_1,\sigma_2,\bar\sigma_2}
\left|{\cal M}_{\sigma_1\bar\sigma_1 : \sigma_2\bar\sigma_2}\right|^2
&=& \left(\frac{gM_W}{c_W^2}\right)^2\left(\frac{g}{c_W}\right)^4\
\frac{k_1^2} {(k_1^2-M_Z^2)^2+M_Z^2\Gamma_Z^2}\
\frac{k_2^2} {(k_2^2-M_Z^2)^2+M_Z^2\Gamma_Z^2} \nonumber \\[3mm]
&\times &
(v_{f_1}^2+a_{f_1}^2) (v_{f_2}^2+a_{f_2}^2)\ \frac{128\pi}{9}\
\sum_{i=1}^9\ C_i\ f_i(\theta_1,\theta_2,\Phi)
\end{eqnarray}
with $\Phi=\phi_1+\phi_2$ and $\eta_i=2v_{f_i}a_{f_i}/(v_{f_i}^2+a_{f_i}^2)$.
The normalized 9 angular distributions are given by
\footnote{Note that
$\int f_i(\theta_1,\theta_2,\Phi) {\rm d}c_{\theta_1}{\rm d}c_{\theta_2}{\rm d}\Phi
=\delta_{i1}+\delta_{i3}$.  }
\begin{eqnarray}
f_1(\theta_1,\theta_2,\Phi)&=&\frac{9}{128\pi}\left[
(1+c_{\theta_1}^2)(1+c_{\theta_2}^2)+4\eta_1\eta_2c_{\theta_1}c_{\theta_2}
\right]\,, \nonumber \\[2mm]
f_2(\theta_1,\theta_2,\Phi)&=&\frac{9}{128\pi}\left\{
-2\left[\eta_1c_{\theta_1}(1+c_{\theta_2}^2)
       +\eta_2c_{\theta_2}(1+c_{\theta_1}^2)
\right] \right\}\,, \nonumber \\[2mm]
f_3(\theta_1,\theta_2,\Phi)&=&\frac{9}{128\pi}\left[
4s_{\theta_1}^2s_{\theta_2}^2
\right]\,, \nonumber \\[2mm]
f_4(\theta_1,\theta_2,\Phi)&=&\frac{9}{128\pi}\left[
4(c_{\theta_1}c_{\theta_2}+\eta_1\eta_2)s_{\theta_1}s_{\theta_2}c_\Phi
\right]\,, \nonumber \\[2mm]
f_5(\theta_1,\theta_2,\Phi)&=&\frac{9}{128\pi}\left[
-4(c_{\theta_1}c_{\theta_2}+\eta_1\eta_2)s_{\theta_1}s_{\theta_2}s_\Phi
\right]\,, \nonumber \\[2mm]
f_6(\theta_1,\theta_2,\Phi)&=&\frac{9}{128\pi}\left[
-4(\eta_1c_{\theta_2}+\eta_2c_{\theta_1})s_{\theta_1}s_{\theta_2}c_\Phi
\right]\,, \nonumber \\[2mm]
f_7(\theta_1,\theta_2,\Phi)&=&\frac{9}{128\pi}\left[
4(\eta_1c_{\theta_2}+\eta_2c_{\theta_1})s_{\theta_1}s_{\theta_2}s_\Phi
\right]\,, \nonumber \\[2mm]
f_8(\theta_1,\theta_2,\Phi)&=&\frac{9}{128\pi}\left[
s_{\theta_1}^2s_{\theta_2}^2c_{2\Phi}
\right]\,, \nonumber \\[2mm]
f_9(\theta_1,\theta_2,\Phi)&=&\frac{9}{128\pi}\left[
-s_{\theta_1}^2s_{\theta_2}^2s_{2\Phi}
\right]\,.
\end{eqnarray}
Also, the 9 angular coefficients $C_{1-9}$, which are
combinations of the reduced helicity amplitudes $\langle+\rangle$, $\langle-\rangle$,
and $\langle0\rangle$, are defined as
\begin{eqnarray}
\label{eq:cs}
C_1&\equiv &
\left|\langle+\rangle\right|^2+\left|\langle-\rangle\right|^2
\,, \ \ \
C_2\equiv 
\left|\langle+\rangle\right|^2-\left|\langle-\rangle\right|^2
\,, \ \ \
C_3\equiv 
\left|\langle0\rangle\right|^2\,,
\nonumber \\[2mm]
C_4&\equiv &
\real\left[\langle + \rangle\langle 0\rangle^*
+\langle-\rangle\langle0\rangle^*\right]
\,, \ \ \
C_5\equiv 
\imag\left[\langle+\rangle\langle0\rangle^*
-\langle-\rangle\langle0\rangle^*\right]\,,
\nonumber \\[2mm]
C_6&\equiv  &
\real\left[\langle+\rangle\langle0\rangle^*
-\langle-\rangle\langle0\rangle^*\right]
\,, \ \ \
C_7\equiv 
\imag\left[\langle+\rangle\langle0\rangle^*
+\langle-\rangle\langle0\rangle^*\right]\,,
\nonumber \\[2mm]
C_8&\equiv &
2\real\left[\langle+\rangle\langle-\rangle^*\right]
\,, \ \ \
C_9\equiv 
2\imag\left[\langle+\rangle\langle-\rangle^*\right]\,.
\end{eqnarray}
Under CP   and CP$\widetilde{\rm T}$~\footnote{
$\widetilde{\rm T}$ denotes the naive  time-reversal transformation
under which the the  matrix element gets complex conjugated.
}   
transformations, the reduced $H$-$Z$-$Z$ helicity  amplitudes  transform as follows:
\begin{equation}
\langle \lambda\rangle
\, \stackrel{\rm CP}{\leftrightarrow}  \,
\langle -\lambda\rangle\,, \qquad
\langle \lambda\rangle
\, \stackrel{\rm CP\widetilde{\rm T}}{\leftrightarrow} \,
\langle -\lambda\rangle^*\,.
\end{equation}
We note that the CP parities of $C_2$, $C_5$ ,$C_6$ and $C_9$ 
are {\it negative} (CP odd) implying that they
are non-vanishing  only when $\{g_{_{HZZ}},S_H^{ZZ}\}$ and $P_H^{ZZ}$
exist simultaneously.
Furthermore, 
the CP$\widetilde{\rm T}$ parities of $C_2$, $C_6$, $C_7$ are 
(CP$\widetilde{\rm T}$ odd), which implies that they can only be induced
by non-vanishing absorptive (or imaginary) parts of $S_H^{ZZ}$ and/or $P_H^{ZZ}$.

\subsection{Angular observables}
The partial decay width of the process $H\to ZZ \to 2\ell_1 2\ell_2$ is given by
\begin{eqnarray}
\label{eq:angdist}
d\Gamma &=& \frac{1}{2 M_H} \left(
\sum_{\sigma_1,\bar\sigma_1,\sigma_2,\bar\sigma_2}
\left|{\cal M}_{\sigma_1\bar\sigma_1 : \sigma_2\bar\sigma_2}\right|^2
\right)\, d\Phi_4
\nonumber \\[2mm]
&=& \frac{1}{2^{13}\, \pi^6\, M_H}\,
\lambda^{1/2}(1,k_1^2/M_H^2,k_2^2/M_H^2)
\sqrt{k_1^2}\sqrt{k_2^2} 
\nonumber \\[2mm]
&&\hspace{2.0cm} \times
\left(
\sum_{\sigma_1,\bar\sigma_1,\sigma_2,\bar\sigma_2}
\left|{\cal M}_{\sigma_1\bar\sigma_1 : \sigma_2\bar\sigma_2}\right|^2
\right)\
d\sqrt{k_1^2}\,d\sqrt{k_2^2}\,dc_{\theta_1}\,dc_{\theta_2}\,d\Phi\,.
\end{eqnarray}
After integrating over $\sqrt{k_1^2}$ and $\sqrt{k_2^2}$, we obtain
\begin{equation}
\frac{1}{\Gamma}\frac{{\rm d}\Gamma}
{ {\rm d}c_{\theta_{1}}{\rm d}c_{\theta_{2}} {\rm d}\Phi } =
\sum_{i=1}^9\ \overline{R}_i f_i(\theta_1,\theta_2,\Phi)
\end{equation}
with the 9 angular observables defined by 
\begin{equation}
\label{eq:ris}
\overline{R}_i \equiv 
\frac{w_i \overline{C}_i}{w_1 \overline{C}_1 + w_3 \overline{C}_3}.
\end{equation}
Note that we have introduced the 9 weight factors $w_i$ in 
the definition of the angular observables $\overline{R}_i$
which are defined by
\begin{equation}
w_i\equiv \frac{{\cal F}_i}{{\cal F}\overline{C}_i}
\end{equation}
where the constant angular coefficients at $Z$ pole are given by
\begin{equation}
\overline{C}_i= C_i(k_1^2=M_Z^2,k_2^2=M_Z^2)
\end{equation}
and the numerical factors by
\begin{eqnarray}
{\cal F}&=&
\int \lambda^{1/2}(1,k_1^2/M_H^2,k_2^2/M_H^2)\sqrt{k_1^2}\sqrt{k_2^2} \
\frac{k_1^2} {(k_1^2-M_Z^2)^2+M_Z^2\Gamma_Z^2}\
\frac{k_2^2} {(k_2^2-M_Z^2)^2+M_Z^2\Gamma_Z^2}\
d\sqrt{k_1^2}\,d\sqrt{k_2^2}\,, \nonumber \\[5mm]
{\cal F}_i&=&
\int \lambda^{1/2}(1,k_1^2/M_H^2,k_2^2/M_H^2)\sqrt{k_1^2}\sqrt{k_2^2} \
C_i(k_1^2,k_2^2)\
\nonumber \\
&&\hspace{3.5cm} \times \
\frac{k_1^2} {(k_1^2-M_Z^2)^2+M_Z^2\Gamma_Z^2}\
\frac{k_2^2} {(k_2^2-M_Z^2)^2+M_Z^2\Gamma_Z^2}\
d\sqrt{k_1^2}\,d\sqrt{k_2^2}\,. \\ \nonumber
\end{eqnarray}
In general, the angular coefficients $C_i$ depends of the momenta of $Z$ bosons.
When $M_H>2 M_Z$, the two decaying $Z$ bosons are predominantly on-shell. 
In this case, one may have $w_i=1$ by
adopting the narrow-width approximation (NWA) for the intermediate
$Z$ bosons.
We therefore note that the deviation of the weight factor from 
unity measures the accuracy of the approximation.

After integrating over any two of the angles $\theta_1$, $\theta_2$, and $\Phi$,
one may obtain the following analytic expressions for the one-dimensional 
angular distributions in terms of the $Z$-pole angular coefficients
$\overline{C}_{1-9}$:
\begin{eqnarray}
\label{eq:ang-analy}
\frac{1}{\Gamma}\frac{{\rm d}\Gamma}{{\rm d}c_{\theta_{1,2}}}
&=&
\frac{3}{8}\overline{R}_1 \left(1+c_{\theta_{1,2}}^2\right)
-\frac{3\eta_{1,2}}{4} \overline{R}_2\ c_{\theta_{1,2} }
+\frac{3}{4}\overline{R}_3 \left(1-c_{\theta_{1,2}}^2\right)\,,
\nonumber \\ [3mm]
\frac{1}{\Gamma}\frac{{\rm d}\Gamma}{{\rm d}\Phi}
&=&
\frac{1}{2\pi}
+\frac{9\pi\eta_1\eta_2}{128}\left(\overline{R}_4\ c_\Phi
- \overline{R}_5\ s_\Phi\right) 
+\frac{1}{8\pi}\left(\overline{R}_8\ c_{2\Phi} - \overline{R}_9\
s_{2\Phi}\right)
\end{eqnarray}
with
\begin{eqnarray}
\label{eq:gamhz4l}
\Gamma &=&\frac{1}{2^63^2\pi^5 M_H}
\left(\frac{gM_W}{c_W^2}\right)^2\left(\frac{g}{c_W}\right)^4
\left(v_{f_1}^2+a_{f_1}^2\right) \
\left(v_{f_2}^2+a_{f_2}^2\right) \
\left(w_1\overline{C}_1 +w_3\overline{C}_3\right)\ {\cal F}\,. 
\\[-5mm] \nonumber
\end{eqnarray}
First, we note that only $C_{1,2,3}$ contribute to the
$c_{\theta_{1,2}}$ distributions.
When $S_H^{ZZ}$ and $P_H^{ZZ}$ are real or when their imaginary parts are negligible,
$C_2=0$ and the linear term is vanishing and
the $c_{\theta_{1,2}}$ distributions are symmetric and parabolic.
The coefficients $C_{4,5}$ and $C_{8,9}$ together with $C_{1,3}$ in
the denominators are contributing to
the $\Phi$ distribution.
For the decay $ZZ\to 4\ell$,
with $\eta_\ell=2v_\ell a_\ell/(v_\ell^2+a_\ell^2)=0.150$ for charged leptons,
$9\pi\eta_\ell^2/128 \sim 0.005$ and
$1/8\pi\sim 0.04$,
the $\Phi$ distribution mostly varies as $s_{2\Phi}$ and $c_{2\Phi}$.
Finally, we note that
the angular observables $\overline{R}_{6,7}$ never appear in
the one-dimensional angular distributions since
$C_{6,7}$ do not contribute to them.
To probe $C_{6,7}$, one may need to study two-dimensional angular distributions such as
$c_{\theta_1}$-$\Phi$ and $c_{\theta_2}$-$\Phi$  distributions.

The angular observables $\overline{R}_{1,2,3}$
can be obtained by the $c_{\theta_{1,2}}$ polynomial fitting to the
$\theta_{1,2}$ distributions, while
 $\overline{R}_{4,5,8,9}$ can be obtained
either 
by the Fourier analysis of the $\Phi$ distribution or by performing 
the fit to the distribution.
We emphasize that it is important to measure all the angular 
observables $\overline{R}_i$
since each of them has different physical implications.
A non-vanishing $\overline{R}_{2}$, for example, may imply the existence of 
New Physics  particles  with mass less than $M_H/2$;
non-vanishing $\overline{R}_{5,9}$ 
may imply that there should be an extra source of CP violation
beyond the Cabibbo--Kobayashi--Maskawa  (CKM) phase in the SM.

The measurements of the angular observables $\overline{R}_i$ alone, however,
cannot determine the absolute size of the couplings 
of $g_{_{HZZ}}$, $S_H^{ZZ}$, and $P_H^{ZZ}$.
For this purpose one may need to measure the quantity
$w_1\overline{C}_1+w_3\overline{C}_3$.
From Eq.~(\ref{eq:gamhz4l}), using ${\cal F}=2280$, we have
\begin{eqnarray}
\Gamma &=& 2.78 \times 10^{-4}\ (w_1\overline{C}_1+w_3\overline{C}_3)\ {\rm GeV}
\nonumber \\
&=&\Gamma^H_{\rm tot}\ B(H\to ZZ \to 2\ell_1 2\ell_2)
\simeq
\Gamma^H_{\rm tot}\ B(H\to ZZ)\ \left[B(Z\to \ell\ell)\right]^2
\end{eqnarray}
where
$\Gamma^H_{\rm tot}$ denotes the total decay width of the heavy Higgs boson $H$.
Assuming information on $B(H\to ZZ)$ can be extracted from $\sigma\cdot B$ 
measurements by
considering several $H$ production and decay processes, and
together with an independent measurement of  the total decay width,
one may determine the combination of $w_1\overline{C}_1+w_3\overline{C}_3$:
\begin{equation}
w_1\overline{C}_1+w_3\overline{C}_3 = 4.1\ \frac{\Gamma^H_{\rm tot}}{\rm GeV}\
B(H\to ZZ)
\end{equation}
where we use $B(Z\to \ell\ell)=3.3658\times 10^{-2}$.

\section{Numerical Analysis}

For numerical analysis we are taking $M_H=260$ GeV.
First, this choice of $M_H$ ensures two on-shell 
$Z$ bosons, and slightly above the $2M_h$ decay threshold,
such that $B(H\to ZZ)$ may be comparable to $B(H\to hh,hZ)$. Simultaneously,
it is far below the $2m_t$ threshold, and so $B(H\to t\bar t)=0$.
Furthermore, the form factors $S_H^{ZZ}$ and $P_H^{ZZ}$
are most likely to be real, 
because, with $M_H<2 m_t$, their imaginary (absorptive) parts are
negligible unless there exist light (lighter than $M_H/2=130$ GeV)
particles which significantly couple to $H$.
This significantly simplifies our numerical analysis and there are
only 3 real parameters to vary.
Incidentally, we note that a heavy scalar with a mass around 270 GeV
may explain some excesses observed in LHC Run I data or 
those observed in 
measurements of the transverse momentum of $h$,
$h$ production associated with top quarks, and 
searches for $hh$ and $VV$ resonances
\cite{vonBuddenbrock:2015ema,vonBuddenbrock:2016rmr}.

Bearing this in mind we consider the following 6 representative
scenarios:
\begin{itemize}
\item {\bf S1} : $\left(g_{_{HZZ}}, S_H^{ZZ},  P_H^{ZZ}\right)  = (0.1, 0, 0)$ 
\item {\bf S2} : $\left(g_{_{HZZ}}, S_H^{ZZ},  P_H^{ZZ}\right)  = (0, 0.1, 0)$ 
\item {\bf S3} : $\left(g_{_{HZZ}}, S_H^{ZZ},  P_H^{ZZ}\right)  = (0, 0, 0.1)$ 
\item {\bf S4} : $\left(g_{_{HZZ}}, S_H^{ZZ},  P_H^{ZZ}\right)  = (0, 0.1, 0.1)$ 
\item {\bf S5} : $\left(g_{_{HZZ}}, S_H^{ZZ},  P_H^{ZZ}\right)  = (0, 0.1, -0.1)$ 
\item {\bf S6} : $\left(g_{_{HZZ}}, S_H^{ZZ},  P_H^{ZZ}\right)  = (0.032, 0.1, 0.1)$ 
\end{itemize}
In the first three scenarios of {\bf S1}, {\bf S2}, and {\bf S3}, only one of
the couplings is non-vanishing and CP is conserved.
In the scenarios of {\bf S4} and {\bf S5}, CP is violated and
the couplings $S_H^{ZZ}$ and $P_H^{ZZ}$ take on opposite relative phases.
In the scenario {\bf S6}, all three couplings are non-zero,
with enhancement of the longitudinal component 
$\langle0\rangle$ of the amplitude for a heavier Higgs boson,
the chosen values for the three couplings contribute 
more or less equally to the amplitude squared: see Eq.~(\ref{eq:HZZ}).
Finally, we found that the weight factors lie between $0.99$ and $1.02$,
and therefore we safely take $w_{1-9}=1$ in our numerical study.

\begin{table}[!t]
\caption{\label{tab:scenarios}
{\it The 6 scenarios considered and the 9 angular coefficients at $Z$ pole.
Note that $\overline{C_2}$, $\overline{C_5}$, $\overline{C_6}$, and $\overline{C_9}$ 
are {\rm CP}-odd indicated by 
their minus($-$) {\rm CP} parities, see the first sign in the square brackets.
And when $S_H^{ZZ}$ and $P_H^{ZZ}$ are real
as taken in our numerical study,
the coefficients $\overline{C_2}$, $\overline{C_6}$ and $\overline{C_7}$ 
are identically vanishing indicated
by their minus($-$) {\rm CP}$\widetilde{\rm T}$ parities,
see the second sign in the square brackets.
}
}
{\footnotesize
\begin{center}
\begin{tabular}{|c|ccc|rrrrrrrrr|}
\hline
& $g_{_{HZZ}}\ $ & $S_H^{ZZ}\ $ & $P_H^{ZZ}\ $ &
$\frac{\overline{C}_1[++]}{10^{-2}}$\  & 
$\frac{\overline{C}_2[--]}{10^{-2}}$\  & 
$\frac{\overline{C}_3[++]}{10^{-2}}$\  &
$\frac{\overline{C}_4[++]}{10^{-2}}$\  & 
$\frac{\overline{C}_5[-+]}{10^{-2}}$\  &
$\frac{\overline{C}_6[--]}{10^{-2}}$\  & 
$\frac{\overline{C}_7[+-]}{10^{-2}}$\  & 
$\frac{\overline{C}_8[++]}{10^{-2}}$\  &
$\frac{\overline{C}_9[-+]}{10^{-2}}$\ 
\\ \hline
{\bf S1} & $0.1$ & $0$ & $0$ &
$2.00$ & $0.00$ & $9.39$ & $-6.13$ & $0.00$ & $0.00$ & $0.00$ & $2.00$ & $0.00$
\\ \hline
{\bf S2} & $0$ & $0.1$ & $0$ &
$1.14$ & $0.00$ & $0.0605$ & $-0.371$ & $0.00$ & $0.00$ & $0.00$ & $1.14$ & $0.00$
\\ \hline
{\bf S3} & $0$ & $0$ & $0.1$ &
$1.02$ & $0.00$ & $0.00$ & $0.00$ & $0.00$ & $0.00$ & $0.00$ & $-1.02$ & $0.00$
\\ \hline
{\bf S4} & $0$ & $0.1$ & $0.1$ &
$2.15$ & $0.00$ & $0.0605$ & $-0.371$ & $0.351$ & $0.00$ & $0.00$ & $0.121$ & $-2.15$
\\ \hline
{\bf S5} & $0$ & $0.1$ & $-0.1~~\,$ &
$2.15$ & $0.00$ & $0.0605$ & $-0.371$ & $-0.351$ & $0.00$ & $0.00$ & $0.121$ & $2.15$
\\ \hline
{\bf S6} & $0.032$ & $0.1$ & $0.1$ &
$1.39$ & $0.00$ & $0.540$ & $0.638$ & $-1.05$ & $0.00$ & $0.00$ & $-0.639$ & $-1.24$
\\ \hline
\end{tabular}
\end{center}
}
\end{table}

\begin{table}
\caption{\label{tab:ratios}
{\it The 6 angular observables 
$\overline{R}_i = \overline{C}_i/(\overline{C}_1+\overline{C}_3)$ 
with $i=1,3,4,5,8,9$
taking $w_{1-9}=1$ and the value of $\overline{C}_1+\overline{C}_3$
for the 6 scenarios under consideration.
The CP and CP$\widetilde{\rm T}$ parities of each observable
are shown in the square brackets.
}
}
{\footnotesize
\begin{center}
\begin{tabular}{|c|ccc|rrrrrr|c|}
\hline
& $g_{_{HZZ}}\ $ & $S_H^{ZZ}\ $ & $P_H^{ZZ}\ $ &
$\ \ \overline{R}_1[++]$ & $\ \ \overline{R}_3[++]$ &
$\ \ \overline{R}_4[++]$ & $\ \ \overline{R}_5[-+]$ &
$\ \ \overline{R}_8[++]$ & $\ \ \overline{R}_9[-+] \ \ $ & 
$\ \ (\overline{C}_1+\overline{C}_3)[++] \times 10^{2} \ \ $ 
\\ \hline
{\bf S1} & $0.1$ & $0$ &  $0$ &
$0.176$ & $0.824$ & $-0.538$ & $0.00$ &  $0.176$ & $0.00\ \ $ &
$11.4$
\\ \hline
{\bf S2} & $0$ & $0.1$ & $0$ &
$0.950$ & $0.0505$ & $-0.310$ & $0.00$ & $0.950$ & $0.00\ \ $ &
$1.20$
\\ \hline
{\bf S3} & $0$ & $0$ & $0.1$ &
$1.00$ & $0.00$ & $0.00$ & $0.00$ &  $-1.00$ & $0.00\ \ $ &
$1.02$
\\ \hline
{\bf S4} & $0$ & $0.1$ & $0.1$ &
$0.973$ &  $0.0273$ & $-0.168$ & $0.158$  & $0.0547$ & $-0.971\ \ $ &
$2.21$
\\ \hline
{\bf S5} & $0$ & $0.1$ & $-0.1~~\,$ &
$0.973$ & $0.0273$ & $-0.168$ & $-0.158$ &  $0.0547$ & $0.971\ \ $ &
$2.21$
\\ \hline
{\bf S6} & $0.032$ & $0.1$ & $0.1$ &
$0.721$ & $0.280$ & $0.330$ & $-0.542$ &  $-0.331$ & $-0.640\ \ $ &
$1.93$
\\ \hline
\end{tabular}
\end{center}
}
\end{table}

\begin{figure}[t!]
\begin{center}
\includegraphics[width=16.3cm]{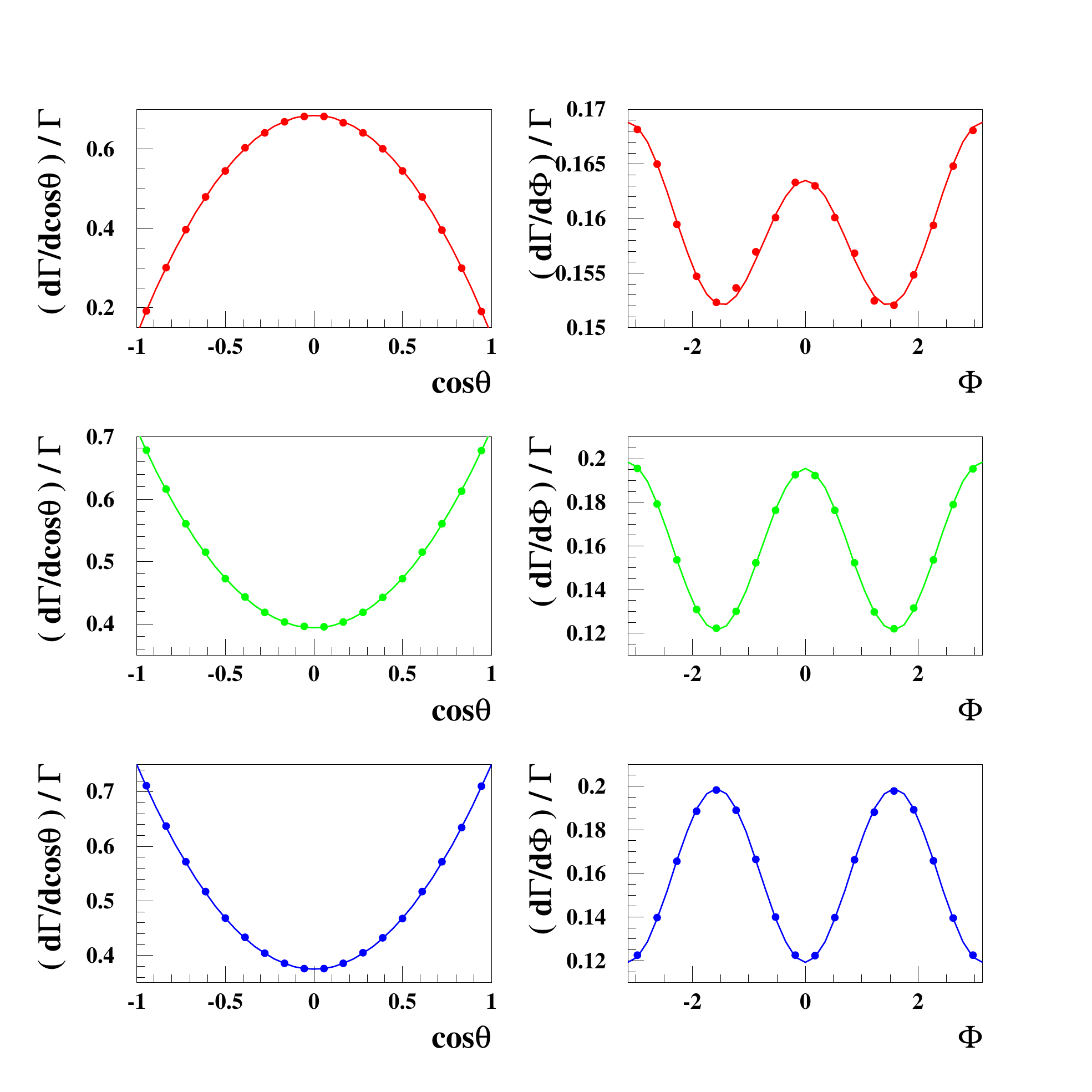}
\end{center}
\vspace{-1.0cm}
\caption{\it
The normalized angular distributions (solid dots)
generated according to the matrix element in Eq.~(\ref{eq:hzz_helamp}) 
with
{\bf S1}:$\left(g_{_{HZZ}},S^{ZZ}_H,P^{ZZ}_H\right)=(0.1,0,0)$ (upper),
{\bf S2}:$\left(g_{_{HZZ}},S^{ZZ}_H,P^{ZZ}_H\right)=(0,0.1,0)$ (middle), and
{\bf S3}:$\left(g_{_{HZZ}},S^{ZZ}_H,P^{ZZ}_H\right)=(0,0,0.1)$ (lower).
The solid lines are drawn using the analytic expressions for
the angular distributions in Eq.~(\ref{eq:ang-analy})
with $w_{1-9}=1$.
}
\label{fig:zz123_a}
\end{figure}

\begin{figure}[t!]
\begin{center}
\includegraphics[width=16.3cm]{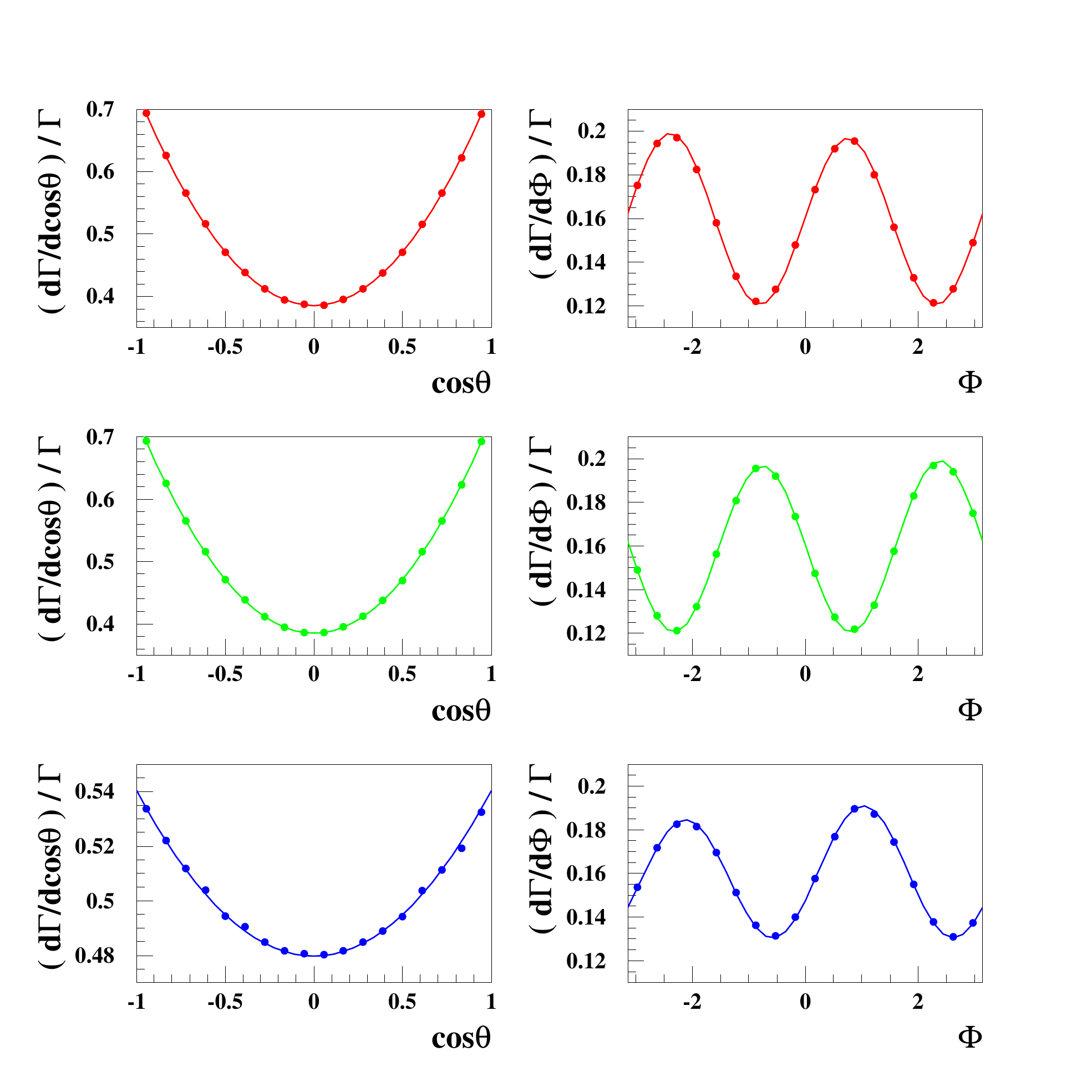}
\end{center}
\vspace{-1.0cm}
\caption{\it
The normalized angular distributions (solid dots)
generated according to the matrix element in Eq.~(\ref{eq:hzz_helamp}) 
with
{\bf S4}:$\left(g_{_{HZZ}},S^{ZZ}_H,P^{ZZ}_H\right)=(0,0.1,0.1)$ (upper),
{\bf S5}:$\left(g_{_{HZZ}},S^{ZZ}_H,P^{ZZ}_H\right)=(0,0.1,-0.1)$ (middle), and
{\bf S6}:$\left(g_{_{HZZ}},S^{ZZ}_H,P^{ZZ}_H\right)=(0.032,0.1,0.1)$ (lower).
The solid lines are drawn using the analytic expressions for
the angular distributions in Eq.~(\ref{eq:ang-analy})
with $w_{1-9}=1$.
}
\label{fig:zz456_a}
\end{figure}

In Table~\ref{tab:scenarios}, we show the 9 angular coefficients 
$\overline{C_1} - \overline{C_9}$ for the 6 scenarios,
together with their CP and CP$\widetilde{\rm T}$ parities 
in the square brackets.
With only the real component in the form factors
$S_H^{ZZ}$ and $P_H^{ZZ}$, 
the coefficients $\overline{C_2}$, $\overline{C_6}$ and 
$\overline{C_7}$ are identically vanishing in all the scenarios, and
$\overline{C_2}$, $\overline{C_5}$, $\overline{C_6}$ and $\overline{C_9}$ 
further vanish in the CP-conserving scenarios of
{\bf S1}, {\bf S2}, and {\bf S3}.
For {\bf S1}, $\overline{C_3}$ is large due to 
the enhancement of the longitudinal component $\langle0\rangle$
of the 
amplitude for a heavier Higgs boson.
Since the longitudinal amplitude $\langle0\rangle=0$
in the {\bf S3} scenario,
only $\overline{C_1}$ and $\overline{C_8}$ take on non-zero values:
see Eq.~(\ref{eq:cs}). 
In the CP-violating scenarios of {\bf S4}, {\bf S5}, and {\bf S6}, 
all the coefficients with plus ($+$)
CP$\widetilde{\rm T}$ parity are non-vanishing. 
Note that with $g_{_{HZZ}}=0$ in {\bf S4} and {\bf S5} ,
the angular coefficient 
$\overline{C_3}=\left|\langle0\rangle\right|^2=4(M_Z/M_H)^4$ is suppressed:
see Eq.~(\ref{eq:HZZ}).
All the non-vanishing
coefficients are comparable in the scenario {\bf S6}.

In Table \ref{tab:ratios}, we show the 6 non-vanishing 
angular observables involved in the
one-dimensional angular distributions 
under the assumption of real $S_H^{ZZ}$ and $P_H^{ZZ}$, 
together with the values of 
$\overline{C_1}+\overline{C_3}$ for the 6 scenarios. The 
first and second signs in the square brackets
again denote the CP and CP$\widetilde{\rm T}$ parities, respectively.
Taking these values
we show the angular distributions
obtained by the analytic expressions Eq.~(\ref{eq:ang-analy}):
see the solid lines
in Figs.~\ref{fig:zz123_a} and \ref{fig:zz456_a}.
For comparisons 
we superimpose the angular distributions generated 
according to Eq.~(\ref{eq:hzz_helamp}) as the solid dots.

In the CP-conserving cases shown in Fig.~\ref{fig:zz123_a}, 
the $\cos\theta_{1,2}$ distribution behaves like
$(1-c_{\theta_{1,2}}^2)$ in scenario {\bf S1} because 
$\overline{R}_1 \ll 2\overline{R}_3$, 
while the distributions behave like $(1+c_{\theta_{1,2}}^2)$ with 
$\overline{R}_1 \gg 2\overline{R}_3$ in scenarios {\bf S2} and {\bf S3}.
The $\Phi$ distributions mostly behave according to 
$\overline{R}_8 c_{2\Phi}$ with the sub-leading contributions from
$\overline{R}_4 c_\Phi$ suppressed by $\eta_\ell^2$: 
see Eq.~(\ref{eq:ang-analy}).
The smaller value at $\Phi=0$ compared to those at $\Phi=\pm\pi$ 
in {\bf S1} (upper right) is due to the negative $\overline{R}_4 c_\Phi$ 
contribution.
Note that they are all symmetric about $\Phi = 0$
without CP violation.

In the CP-violating scenarios of {\bf S4} and {\bf S5},
the $\cos\theta_{1,2}$ distribution behaves like 
$(1+c_{\theta_{1,2}}^2)$ with $\overline{R}_1 \gg 2\overline{R}_3$:
see the upper left and middle left frames of Fig.~\ref{fig:zz456_a}.
While in {\bf S6} with $\overline{R}_1$ slightly larger
than $2\overline{R}_3$,
it still behaves as $(1+c_{\theta_{1,2}}^2)$ but its variation
is much smaller compared to the {\bf S4} and {\bf S5} scenarios due to the 
cancellation between the $\overline{R}_1$ and $\overline{R}_3$ terms.
The $\Phi$ distributions mostly behave according to 
$\overline{R}_8 c_{2\Phi}-\overline{R}_9 s_{2\Phi}$
with the sub-leading contributions from
$\overline{R}_4 c_\Phi-\overline{R}_5 s_\Phi$.
We observe that they are no longer symmetric about $\Phi=0$ due to
non-trivial phase shift 
induced by the CP violating terms of $s_{2\Phi}$ and $s_\Phi$.

We observe the complete agreement between the angular distributions
obtained by the analytic expressions in Eq.~(\ref{eq:ang-analy})
and those generated 
according to the helicity amplitude Eq.~(\ref{eq:hzz_helamp}),
and therefore conclude that our analytic expressions provide an 
excellent framework to extract the couplings 
$g_{_{HZZ}}$, $S_H^{ZZ}$, and $P_H^{ZZ}$
and completely measure the properties of a CP-mixed scalar boson $H$
through the angular distributions.

Now we are going to illustrate how well one can 
measure the properties of the $260$ GeV Higgs by taking the example of
scenario {\bf S6} with $\left(g_{_{HZZ}},S^{ZZ}_H,P^{ZZ}_H\right)=(0.032,0.1,0.1)$,
in which all three couplings play almost equal roles.
For this purposes we generate a pseudo dataset with the number of events
$N_{\rm evt}=10^3$ in the range of 
$\sqrt{k_{1,2}^2} = M_Z \pm 4$ GeV by noting that the current upper limit on
$\sigma(gg\to H)\cdot B(H\to ZZ) \lsim 0.1$ pb 
for a 260 GeV Higgs boson at 95 \% C.L.~\cite{h_zz,atlas17}: 
$$
\sigma(gg\to H)\cdot B(H\to ZZ)\cdot 4[B(Z\to \ell\ell)]^2\cdot \epsilon_{4\ell}
\cdot {\cal L} \simeq 10^3
$$
where we naively take the 4-lepton efficiency $\epsilon_{4\ell} \sim 1$ 
\footnote{We find that $\epsilon_{4\ell} \sim (0.95)^4$
by requiring $p_T>25 (5)$ GeV for the leading (sub-leading) lepton
with the rapidity cut $|\eta_\ell| < 2.5$.}
and assume the HL-LHC with the luminosity of
${\cal L}=3/{\rm ab}$.  
Further, we assume the angular resolutions of 
$\Delta\cos\theta=0.1$ and $\Delta\Phi=0.1\pi$.

\begin{figure}[t!]
\begin{center}
\includegraphics[width=8.0cm]{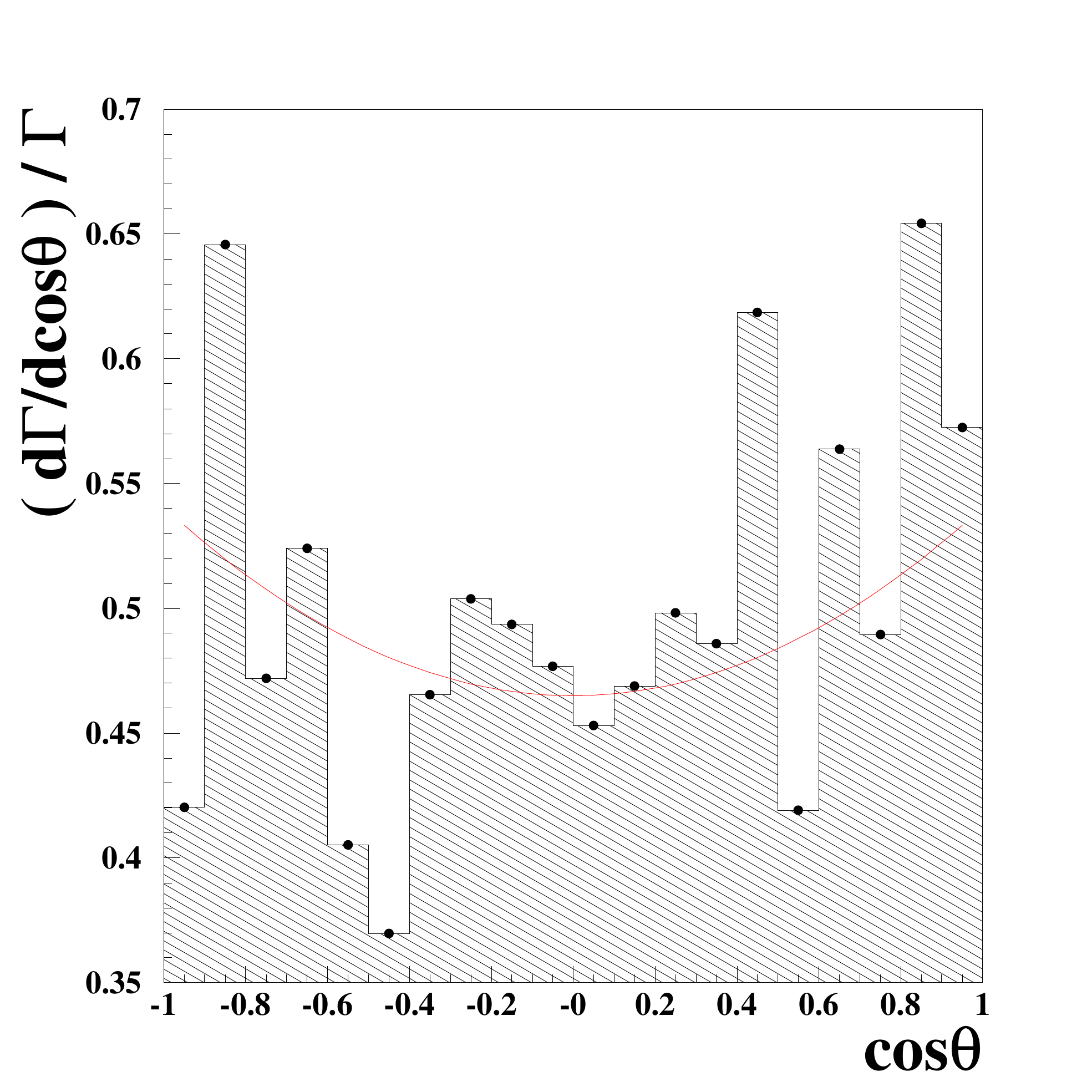}
\includegraphics[width=8.0cm]{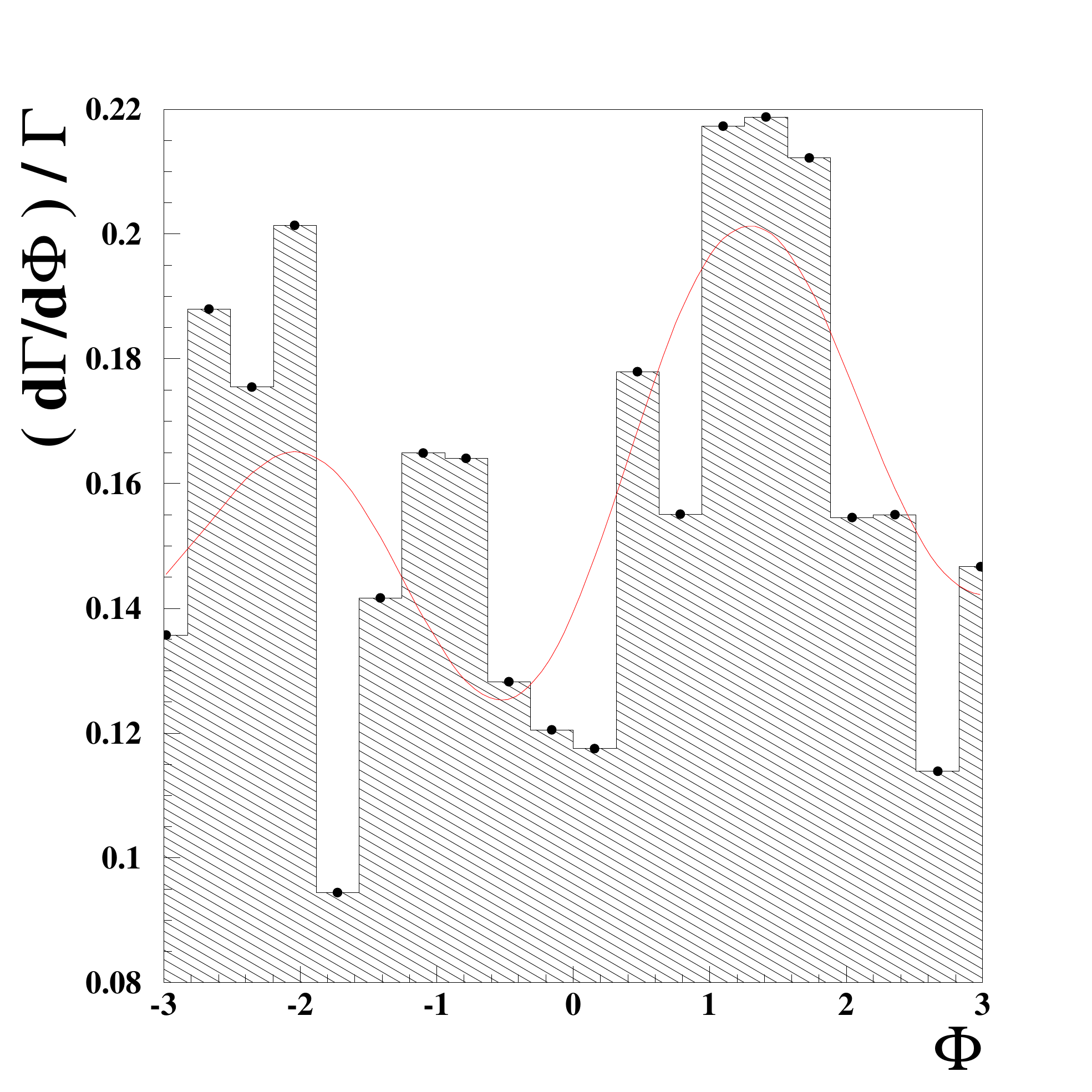}
\end{center}
\vspace{-1.0cm}
\caption{\it {\bf S6}: 
The angular distributions from the pseudo dataset 
of $N_{\rm evt}=10^3$ events generated
with $\sqrt{k_{1,2}^2} = M_Z \pm 4\ {\rm GeV}$, 
$\Delta\cos\theta=0.1$ and
$\Delta\Phi=0.1\pi$.
The results of fitting to the angular distributions with
Eq.~(\ref{eq:ang-analy}) are shown in the (red) solid lines.
}
\label{fig:s6}
\end{figure}

In Fig.~\ref{fig:s6}, the histograms 
show the normalized $\cos\theta$ (left) and $\Phi$ (right) distributions
from the pseudo dataset of $N_{\rm evt}=10^3$ events.
Here the $\cos\theta$ distribution is
the combination of the $\cos\theta_1$ and $\cos\theta_2$ distributions.
One can obtain the angular observables
$\overline{R}_{1,3}$ by fitting to the $\cos\theta$ distribution
with the analytic expression for the 
$1/\Gamma\ {\rm d}\Gamma/{\rm d}c_{\theta_{1,2}}$ in
Eq.~(\ref{eq:ang-analy}). Note we have fixed $\overline{R}_2=0$ in the fitting.
We have found the strong correlation between the $\overline{R}_1$ 
and $\overline{R}_3$
observables with the correlation coefficient $\rho=-0.813$.
The angular observables $\overline{R}_{4,5,8,9}$ can be obtained
by the Fourier analysis of the $\Phi$ distribution. Explicitly, one may have
\begin{eqnarray}
\overline{R}_4 &=& \frac{128}{9\pi^2 \eta_\ell^2}
\int c_\Phi\ \left(\frac{1}{\Gamma}\frac{{\rm d}\Gamma}{{\rm d}\Phi}\right)  
{\rm d}\Phi \,, \ \ \
\overline{R}_5  =  -\ \frac{128}{9\pi^2 \eta_\ell^2}
\int s_\Phi \left(\frac{1}{\Gamma}\frac{{\rm d}\Gamma}{{\rm d}\Phi}\right)
{\rm d}\Phi \,, 
\nonumber \\[2mm]
\overline{R}_8 &=& 8
\int c_{2\Phi}\ \left(\frac{1}{\Gamma}\frac{{\rm d}\Gamma}{{\rm d}\Phi}\right)
{\rm d}\Phi \,, 
\hspace{1.15cm}
\overline{R}_9  =  -8
\int s_{2\Phi}\ \left(\frac{1}{\Gamma}\frac{{\rm d}\Gamma}{{\rm d}\Phi}\right)
{\rm d}\Phi \,.
\\ \nonumber
\end{eqnarray}
The angular observables $\overline{R}_{4,5,8,9}$ can also be obtained
by performing a fit to the $\Phi$ histogram distribution 
with the analytic expression for the $1/\Gamma\ {\rm d}\Gamma/{\rm d}\Phi$ in
Eq.~(\ref{eq:ang-analy}).
We have checked that $\overline{R}_{4,5,8,9}$ from the Fourier analysis
and those from the fitting are 
consistent within errors
\footnote{The output central values obtained from the Fourier analysis
are: $\overline{R}_4=-0.557$, $\overline{R}_5=-3.36$, $\overline{R}_8=-0.543$,
$\overline{R}_9=-0.440$.}. 
In our numerical analysis, we use
the fitted angular observables. 
The results of the fittings are represented by the (red) solid lines in
In Fig.~\ref{fig:s6}.

\begin{table}[!t]
\caption{\label{tab:s6}
{\it The input and output values of the $6$ angular observables
$\overline{R}_{1,3,4,5,8,9}$  
involved in the
one-dimensional angular distributions 
under the assumption of real $S_H^{ZZ}$ and $P_H^{ZZ}$.
We have taken the scenario {\bf S6}:
$\left(g_{_{HZZ}},S^{ZZ}_H,P^{ZZ}_H\right)=(0.032,0.1,0.1)$.
The input values are the same as in Table \ref{tab:ratios}.
The output values have been
obtained by fitting to the $\cos\theta_{1,2}$ and $\Phi$ distributions
in Fig.~\ref{fig:s6}.
The correlation for $\overline{R}_1$ and $\overline{R}_3$
is $\rho=-0.813$, while the correlations among others
are negligible.
For $\overline{C}_1+\overline{C}_3$, we simply assume 20 \% error.
}
}
{\footnotesize
\begin{center}
\begin{tabular}{|c|rrrrrr|c|}
\hline
{\bf S6} &
$\ \ \overline{R}_1[++]$ & $\ \ \overline{R}_3[++]$ &
$\ \ \overline{R}_4[++]$ & $\ \ \overline{R}_5[-+]$ &
$\ \ \overline{R}_8[++]$   & $\ \ \overline{R}_9[-+]\ \ $ & 
$\ \ (\overline{C}_1+\overline{C}_3)[++] \times 10^{2} \ \ $
\\ \hline
Input & 
$0.721$ & $0.280$ & $0.330$ & $-0.542$ & $-0.331$ & $-0.640\ \ $ &
$1.93$ \\
\hline
Output (center value) & 
$0.721$  & 
$0.260$  &
$-0.339$  & 
$-4.07$ & 
$-0.452$ & 
$-0.387  \ \ $ &
$1.93$ \\
Output (parabolic error) & 
$\pm 0.037$ & 
$\pm 0.034$ &
$\pm 1.37$ &   
$\pm 1.45$ &  
$\pm 0.17$ &  
$\pm 0.18\ \ $ &
$\pm 0.386$
\\ \hline
\end{tabular}
\end{center}
}
\end{table}

The details of the fitting results are summarized in
Table \ref{tab:s6} as the output central values together with
the corresponding parabolic errors. 
We observe that the output central values are 
within the  $1$- or $2$-$\sigma$ ranges of the input values.
Note that the CP violation is observed at the $2$-$\sigma$ level
with $\overline{R}_9=-0.387\pm 0.18$.
The observation through another CP-violating observable
$\overline{R}_5$ is also at the $2$-$\sigma$ level:
$\overline{R}_5=-4.07 \pm 1.45$. 
First, the error is $8$ times larger than that of $\overline{R}_9$
because of the $\eta_\ell^2$ suppression factor, 
see Eq.~(\ref{eq:ang-analy}).
Second, this is due to the statistical fluctuation.
We have verified that the central values of 
the observable $\overline{R}_5$ 
are quite close to the input value $-0.542$
if we generate more pseudo datasets of $10^3$ events.

\begin{figure}[t!]
\begin{center}
\includegraphics[width=16.0cm]{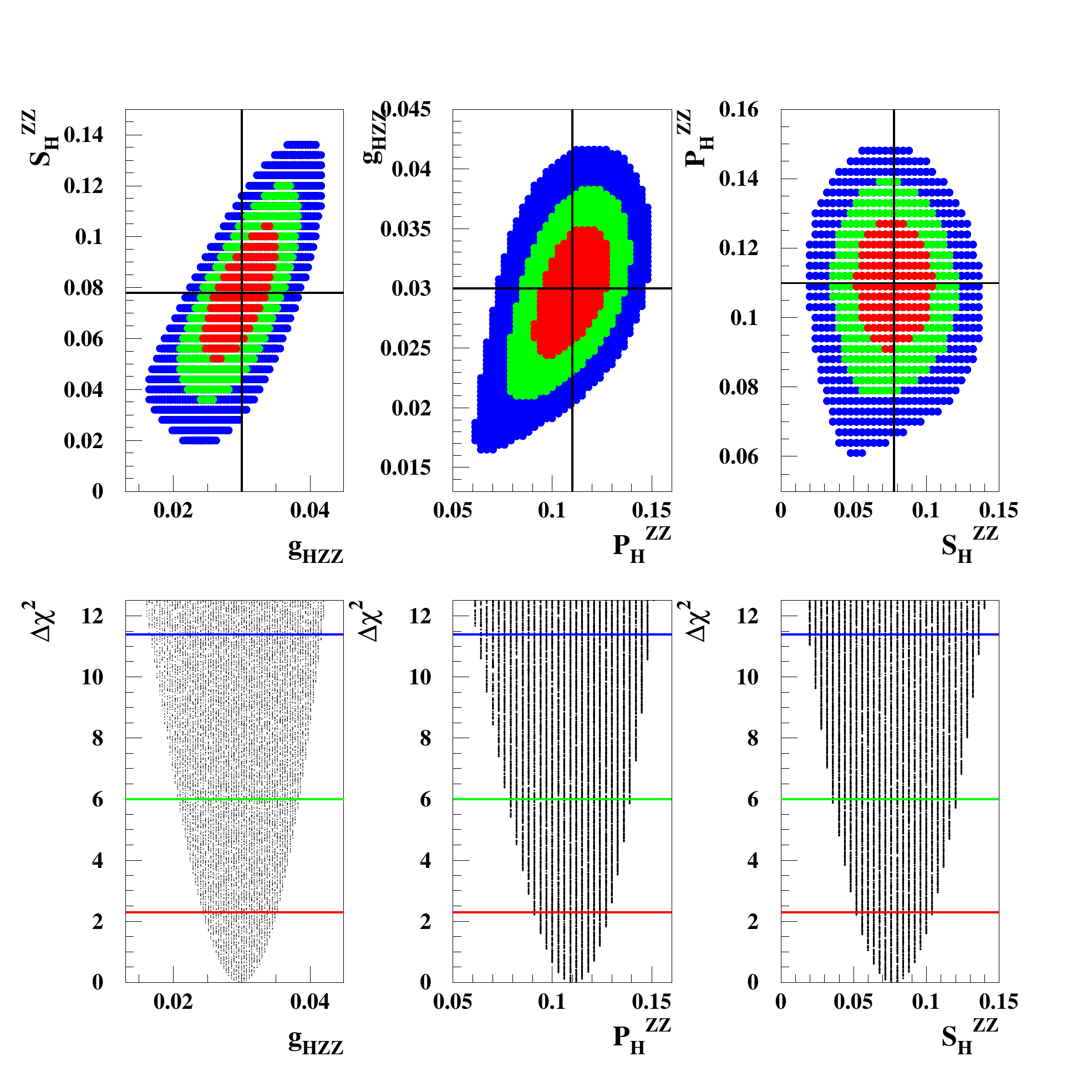}
\end{center}
\vspace{-1.0cm}
\caption{\it 
Upper: The confidence-level (CL) regions for scenario {\bf S6} 
$\left(g_{_{HZZ}},S^{ZZ}_H,P^{ZZ}_H\right)=(0.032,0.1,0.1)$
with $\Delta \chi^2 = 2.3$ (red), $5.99$ (green), and $11.83$ (blue)
above the minimum, which
correspond to confidence levels of
$68.3\%$, $95\%$, and $99.7\%$, respectively.
The vertical and horizontal lines show the best-fit values of
$\left(g_{_{HZZ}},S^{ZZ}_H,P^{ZZ}_H\right)^{\rm best-fit}=(0.030,0.078,0.11)$.
Lower: The scatter plots for
$\Delta \chi^2$ versus $g_{_{HZZ}}$ (left),
$\Delta \chi^2$ versus $P_H^{ZZ}$ (middle), and
$\Delta \chi^2$ versus $S_H^{ZZ}$ (left).
The horizontal lines are for the $68.3\%$ (red), 
$95\%$ (green), and $99.7\%$ (blue) CL regions.
}
\label{fig:cls.s6}
\end{figure}

Now we are ready to carry out our ultimate target
to extract the couplings $g_{_{HZZ}}$, $S_H^{ZZ}$, and $P_H^{ZZ}$
from the 7 observables 
$\overline{R}_{1,3,4,5,8,9}$ and $\overline{C}_1+\overline{C}_3$
by implementing a $\chi^2$ analysis.
We have taken into account the correlation between
$\overline{R}_1$ and $\overline{R}_3$, by using
\begin{eqnarray}
\chi^2(\overline{R}_1;\overline{R}_3) &=& 
\Bigg\{
\frac{\left(\overline{R}_1^{\rm TH}-\overline{R}_1^{\rm EXP}\right)^2}
{\left(\sigma_1^{\rm EXP}\right)^2} +
\frac{\left(\overline{R}_3^{\rm TH}-\overline{R}_3^{\rm EXP}\right)^2}
{\left(\sigma_3^{\rm EXP}\right)^2} 
\nonumber \\
&& \hspace{3.0cm}
-2\rho
\frac{\left(\overline{R}_1^{\rm TH}-\overline{R}_1^{\rm EXP}\right)}
{\sigma_1^{\rm EXP}}
\frac{\left(\overline{R}_3^{\rm TH}-\overline{R}_3^{\rm EXP}\right)}
{\sigma_3^{\rm EXP}}\Bigg\} \Bigg/(1-\rho^2)
\end{eqnarray}
where we calculate $\overline{R}_{1,3}^{\rm TH}$ by 
varying the three couplings $g_{_{HZZ}}$, $S_H^{ZZ}$, and $P_H^{ZZ}$: see
Eqs.~(\ref{eq:HZZ}), (\ref{eq:cs}), and (\ref{eq:ris}).
For $\overline{R}_{1,3}^{\rm EXP}$ and $\sigma_{1,3}^{\rm EXP}$, we have taken
the corresponding central output values and errors shown in Table \ref{tab:s6}.
The $\chi^2$'s for the remaining uncorrelated observables are similarly
calculated and summed. 

In the upper frames of Fig.~\ref{fig:cls.s6},
we show the confidence-level regions of
the $\chi^2$ analysis by varying
$g_{_{HZZ}}$, $S_H^{ZZ}$, and $P_H^{ZZ}$.
We have found that 
$\chi^2_{\rm min}/d.o.f=7.34/(7-3)=1.83$
and the minimum occurs at
\footnote{
Incidentally,
the the angular observables $\overline{R}_{1,3,4,5,8,9}$
and the quantity $\overline{C}_1+\overline{C}_3$ calculated using the
best-fit coupling values are:
$\overline{R}_1^{\rm best-fit}= 0.730$,
$\overline{R}_3^{\rm best-fit}= 0.270$,
$\overline{R}_4^{\rm best-fit}= 0.213$,
$\overline{R}_5^{\rm best-fit}= -0.590$,
$\overline{R}_8^{\rm best-fit}= -0.562$,
$\overline{R}_9^{\rm best-fit}= -0.467$, and
$(\overline{C}_1+\overline{C}_3)^{\rm best-fit}= 1.87\times 10^{-2}$.
Note especially that the value of $\overline{R}_5^{\rm best-fit}$ is 
very close 
to its input value $-0.542$. 
We observe one may infer that the fitted value 
$-4.07$ shown in Table~\ref{tab:s6} could be due to statistical 
fluctuation by comparing it to $\overline{R}_5^{\rm best-fit}$. 
}
\begin{eqnarray}
\label{eq:best}
g_{_{HZZ}}=0.030\pm 0.0035\,; \ \ \
S_H^{ZZ}=0.078\pm 0.017 \,; \ \ \
P_H^{ZZ}=0.11\pm 0.013 \,,
\end{eqnarray}
which are consistent with the input values 
$(0.032,0.1,0.1)$ within $\sim 1$-$\sigma$ ranges.
Therefore, we conclude that the three couplings of $H$ to a $Z$ 
boson pair can be determined with about 
12-20\% errors when $N_{\rm evt}=10^3$.
%
We have  implemented the similar analysis with
$N_{\rm evt} = 10^2$ and found that the couplings can be determined with about
30\% errors.

\section{Conclusions}

We have performed a comprehensive study of the most general couplings
of a spin-0 heavy Higgs boson to a pair of $Z$ bosons up to 
dimension-6 operators, using the angular distributions
in the decay $H\to ZZ \to \ell^+ \ell^- \ell^+ \ell^-$.  Based on the
helicity amplitude method,
we figure out there are 9 types of angular observables $\overline{R}_i\, (i=1-9)$ 
according to their CP and CP$\widetilde{\rm T}$ parities:
four of them ($\overline{R}_{2,5,6,9}$)
are CP odd and three of them ($\overline{R}_{2,6,7}$) CP$\widetilde{\rm T}$ odd.
Furthermore, we find that, among the 9 observables, the 2 
CP$\widetilde{\rm T}$-odd observables 
of $\overline{R}_{6,7}$ are not accessible through one-dimensional 
angular distributions.
We have shown that a certain subset of the 9 angular observables 
can be extracted from one- and two-dimensional angular distributions 
of the four final-state charged leptons
depending on the assumption on $S^{ZZ}_{H}$ and $P^{ZZ}_H$.
The parameters $g_{HZZ}, S^{ZZ}_{H}, P^{ZZ}_H$  can then be determined from 
$\overline{R}_i$'s.
This is our novel strategy for analyzing 
the decay $H\to ZZ \to \ell^+ \ell^- \ell^+ \ell^-$
to measure the properties of a heavy Higgs boson $H$.

We have illustrated with $10^3$ events for $H\to ZZ \to 4\ell$ that
the parameters $g_{HZZ}, S^{ZZ}_{H}, P^{ZZ}_H$
can be determined with only 12-20\% uncertainties
through the one-dimensional $\cos\theta_{1,2}$ and $\Phi$ 
distributions
under the assumption of real $S^{ZZ}_{H}$ and $P^{ZZ}_H$.
This is the major numerical result of this work.

We note that following Eq.~(\ref{eq:ang-analy})
the contributions from the coefficients $C_{4,5}$ 
to the $\Phi$ distribution are suppressed by the factor 
$(9\pi^2/16)\eta_\ell^2$  for the decay $ZZ\to 4\ell$,
because the vector coupling $v_\ell \approx 0.02$ for charged leptons.
On the other hand, if we choose the decay $ZZ\to 4b$,
the contributions from the coefficients $C_{8,9}$
are suppressed by the numerical factor in front of the term
while the contribution from the coefficients $C_{4,5}$ becomes large
because $\eta_b \simeq 0.936$,
and so the $\Phi$ distribution mostly varies as $s_{\Phi}$ and $c_{\Phi}$.
In the case of $ZZ\to 2b2l$, all 4 coefficients of $C_{4,5,8,9}$ 
contribute more or less equally.
This interesting possibility will be explored in a future 
publication~\cite{FUTURE}.

We offer the following further comment in our findings.

\begin{enumerate}

\item 
In principle, the form factors $S^{ZZ}_H$ and $P^{ZZ}_H$ can be complex
when the particles running in the loop are on-shell, e.g, when
$M_H > 2 m_t$, the absorptive part appears. In such a case, the 
CP$\widetilde{\rm T}$
angular observables $\overline{R}_{2,6,7}$ are non-vanishing.
In this case, the two-dimensional 
$c_{\theta_1}$-$\Phi$ and $c_{\theta_2}$-$\Phi$ distributions
may provide information on $\overline{R}_{6,7}$ specifically.


\end{enumerate}

{\it Note added:} At the last stage of this work, we became aware of
a paper~\cite{atlas17} from ATLAS on 
search for heavy $ZZ$ resonances in the $\ell^+ \ell^- \ell^+ \ell^-$ 
and $\ell^+\ell^-\nu\bar\nu$ final states
in which, using data at $\sqrt{s}=13$ TeV with the integrated
luminosity of 36.1/fb, they report observation of
two excesses for $m_{4\ell}$ around 240 and 700 GeV,
each with a local significance of $3.6\ \sigma$.
Especially, the resonance around 240 GeV corresponds to more than 30 events
which may lead to about 3000 events at
the HL-LHC with the luminosity of $3/{\rm ab}$, assumed in this work.
In this case, we note that 
the couplings $g_{HZZ}, S^{ZZ}_{H}, P^{ZZ}_H$
can be determined with about 10\% uncertainties.

\section*{Acknowledgment}  
We thank Bruce Mellado Garcia for helpful discussions  and valuable comments.
This work was supported by
the National Research Foundation of Korea (NRF) grant
No. NRF-2016R1E1A1A01943297.
K.C. was supported by the MoST of Taiwan under grant number 
MOST-105-2112-M-007-028-MY3.
%

\section*{Appendix}
\def\theequation{\Alph{section}.\arabic{equation}}
\begin{appendix}

\setcounter{equation}{0}
\section{The four-body phase space}
Four-body phase space can be factorized into 
\begin{eqnarray}
d\Phi_4(Q\to k_1 k_2\to p_1 \bar p_1 p_2\bar p_2) & = &  
 d\Phi_2 ( Q \to k_1 k_2 ) \times d\Phi_2 (k_1 \to p_1 \bar p_1) \times
 d\Phi_2 ( k_2 \to p_2 \bar p_2) \times  \frac{dk_1^2}{2\pi}\
\frac{dk_2^2}{2\pi}  \nonumber \\
%
&=&
\frac{dk_1^2}{2\pi}\
\frac{dk_2^2}{2\pi}\
\frac{\lambda^{1/2}(1,k_1^2/s,k_2^2/s)}{32\pi^2}
d\cos\Theta^*d\Phi^*\ \nonumber \\[3mm]
&\times & \frac{d\cos\theta_1d\phi_1}{32\pi^2} \
\frac{d\cos\theta_2d\phi_2}{32\pi^2} \
\end{eqnarray}
where $s=Q^2$. For our purpose, we may be able to take
\begin{eqnarray}
d\Phi_4(Q\to k_1 k_2\to p_1 \bar p_1 p_2\bar p_2)&=&
\frac{dk_1^2}{2\pi}\
\frac{dk_2^2}{2\pi}\
\frac{\lambda^{1/2}(1,k_1^2/s,k_2^2/s)}{8\pi}
\frac{d\cos\theta_1d\Phi}{32\pi^2} \
\frac{d\cos\theta_2}{16\pi} \
\end{eqnarray}
\end{appendix}

\clearpage


\end{document}